\documentclass[twocolumn,showpacs,preprintnumbers,amsmath,amssymb]{revtex4}

\usepackage{graphicx}
\usepackage{dcolumn}
\usepackage{bm}

\begin{document}

\title{The eigenspectra of Indian musical drums }

\author{G. Sathej}
 \affiliation{Department of Biotechnology, Indian Institute of Technology - Madras, Chennai 600036, India}
\author{R. Adhikari}%
\affiliation{ The Institute of Mathematical Sciences, CIT Campus, Tharamani, Chennai 600113, India}

\date{\today}

\begin{abstract}
In a family of drums used in the Indian subcontinent, the circular drum head is made of material of non-uniform density. Remarkably, and in contrast to a circular membrane of uniform density, the low eigenmodes of the non-uniform membrane are harmonic. In this work we model the drum head by a non-uniform membrane whose density varies smoothly between two prescribed values. Using a Fourier-Chebyshev spectral collocation method we obtain the eigenmodes and eigenvalues of the drum head. For a suitable choice of parameters, which we find by optimising a cost function, the eigenspectra obtained from our model are in excellent agreement with experimental values. Our model and the numerical method should find application in numerical sound synthesis. 
\end{abstract}

\maketitle

\section{Introduction}

The eigenvalue problems for a string and an uniform circular membrane are classical problems in mathematical physics. The eigenvalues of a string are determined by the zeros of the sine function and so form a harmonic series. The large number of harmonic overtones give the vibrations of a string its musicality. The eigenvalues  of an uniform membrane, on the other hand, are determined by the zeros of Bessel functions. The overtones are not integer multiples of the fundamental. Consequently, the vibrations do not have a strong sense of pitch and, therefore, lack the musicality of string vibrations. 

Several musical traditions have devised means of restoring musicality to the vibrations of circular drums. The Western tympani achieves this by coupling the vibrations of the membrane with the large mass of air enclosed in the kettle below the drum head. For a judicious choice of modes, the combined membrane-air system has harmonic vibrations\cite{rossbook}. A different strategy is used in a whole family of drums used in the Indian subcontinent, where harmonic overtones are obtained by loading the central part of the membrane with material of heavier density.  These drums have a strong sense of pitch, and in performance, are  tuned to match the tonic of the vocalist or the instrumentalist. 

\begin{figure}
\includegraphics[width = 8cm]{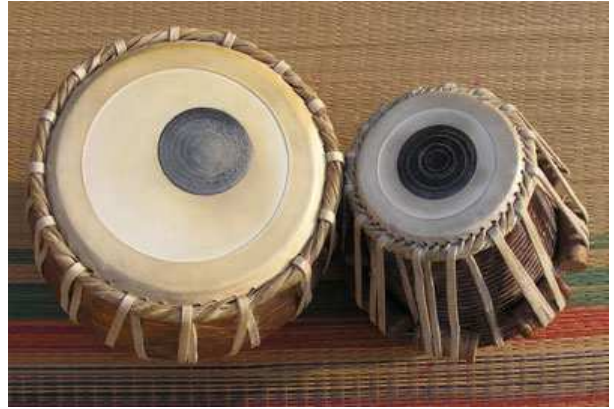}
\caption{\label{fig:tablatopview} The {\it tabla} is a pair of drums consisting of the the left drum, {\it bayan} and the right drum,  {\it dayan}. The cords running along the length of the drums are used to adjust the tension in the membrane, allowing the tonic of the drum to be raised or lowered.} 
\end{figure}

\begin{figure}
\includegraphics[width = 8cm]{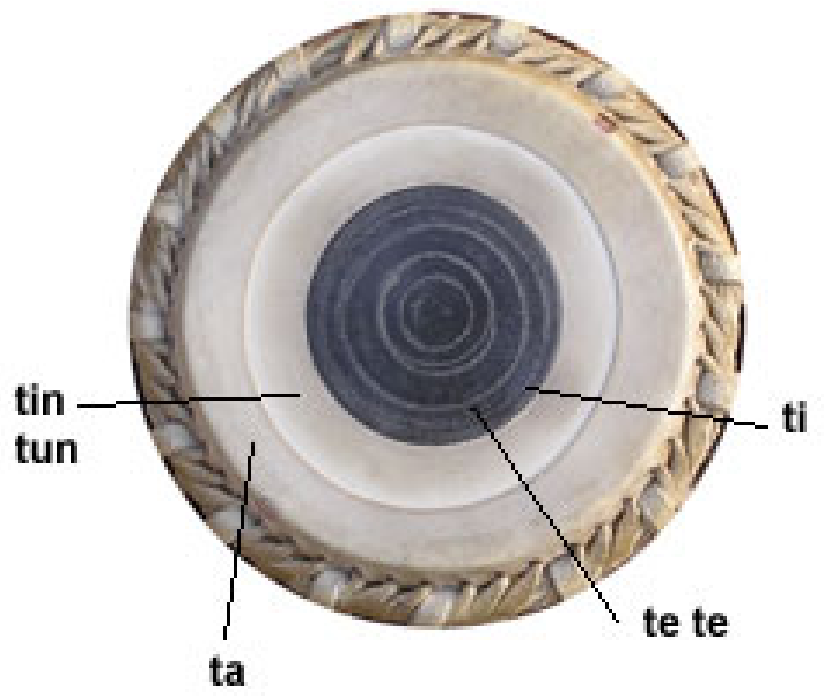}
\includegraphics[width = 6cm]{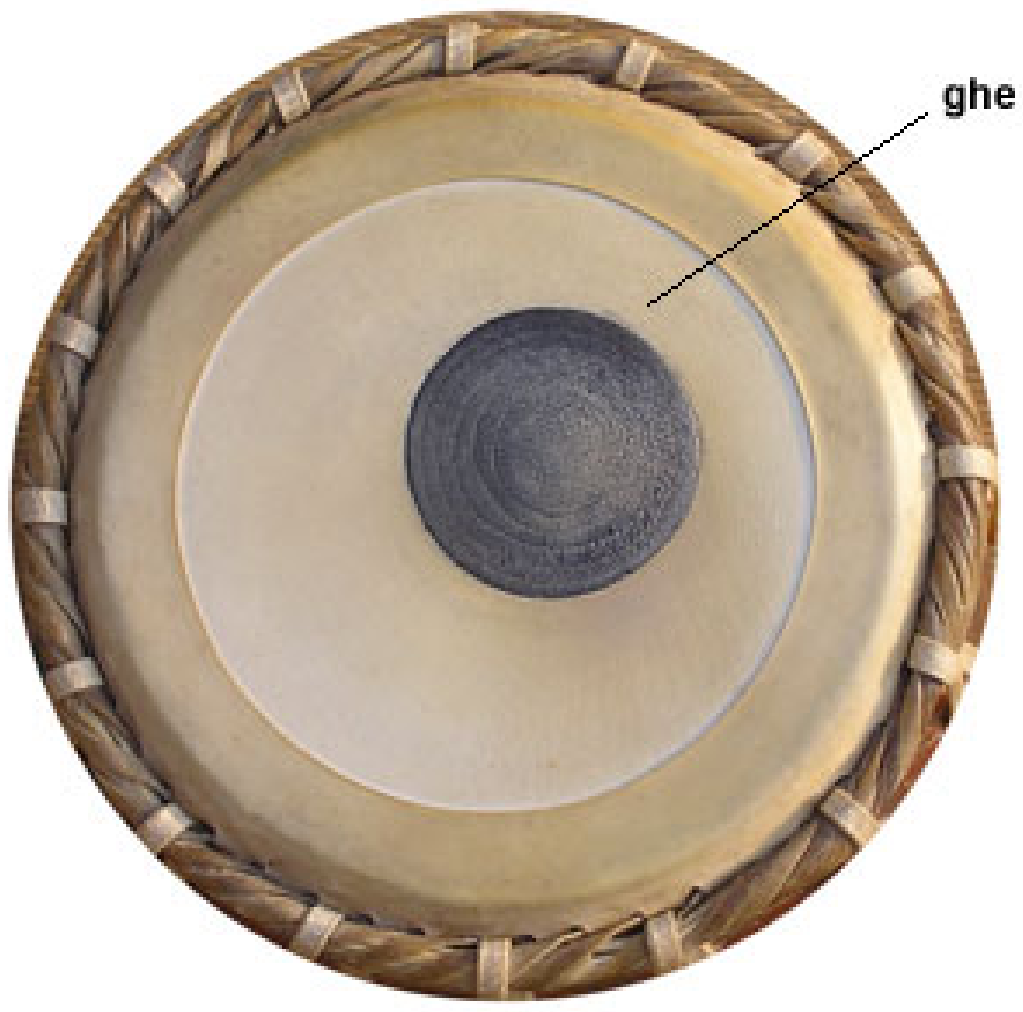}
\caption{\label{fig:dayanbayan} The drumhead of the {\it dayan} (above). The drumhead is made with goatskin and loaded at the center. The central loading patch, the {\it sihai}, is made of a paste of soot, iron filings and flour and applied layer by layer to the goatskin membrane. The {\it sihai} is cracked with a heavy stone after the paste dries, to reduce the rigidity of the material. The thin flap at the outer edge of the drumhead, the {\it kinar}, serves to damp the higher harmonics. Distinct sounds, indicated by the mnemonic syllables {\it tin, tun, ta, ti, tete}, are produced when the drum is struck at different parts. The {\it bayan} (below) is of similar construction, but crucially, the {\it sihai} is placed eccentrically on the membrane. The {\it bayan} has a smaller range of sounds, the principal one being the mnemonic syllable {\it ghe}. The pitch of this syllable is modulated by changing the position of the heel of the hand along the broadest part of the unloaded region of the drum head}.
\end{figure}

The two most popular drums of this family are the South Indian {\it mridangam} and the North Indian {\it tabla}. The {\it mridangam} is a single drum covered on both sides with drum heads made of leather, while the tabla is a pair of drums, the {\it dayan} and the {\it bayan}, each of which have a single drum head (Fig.~\ref{fig:tablatopview}). The loading in the dayan is concentric to the membrane, while in the bayan the loading is eccentric(Fig.~\ref{fig:dayanbayan}). 

Raman made the first scientific study of this family of drums.\cite{ram34} In a series of experiments, Raman and coworkers obtained the eigenmodes and eigenvalues of the {\it mridangam}, showing that the first nine normal modes gave five very nearly harmonic tones. The higher overtones were noticably anharmonic, but Raman noted features in the construction of design to supress the higher overtones.  Subsequently, Ramakrishna and Sondhi\cite{rs54} modelled the drum head as a composite membrane of two distinct densities, with the caveat that ``the density of the loaded region is not constant ... but decreases gradually''. With this simplification, and for concentric loading, the eigenvalue problem could be solved analytically in terms of Bessel and trigonometric functions. The eigenvalues of the composite membrane model agree with Raman's experimental values to within 10\%. Solving the composite membrane model for eccentric loading is considerably more difficult due to lack of circular symmetry. An exact solution for the eigenmodes in terms of known functions is not available. Two approximate solutions\cite{rs57,rah58} have been presented, but the agreement with experimental values is generally poor. Little, therefore, is known about the eigenspectrum of the eccentrically loaded drum head. 

The purpose of this work is two-fold. The first is to present a mathematical model for the loaded drum head of the Indian musical drums, using the {\it tabla} as the prototypical example. The second is to present a high-resolution numerical method, based on Fourier-Chebyshev collocation,  which may be used to obtain the eigenvalues and eigenmodes of a non-uniform circular membrane with an arbitrary variation of the mass density.  Using the numerical method we obtain the eigenspectrum of our model drumhead, for both concentric and eccentric loadings. For concentric loading, our results are in excellent agreement with Raman's experimental values and offer an improvement over the composite membrane model of  Ramakrishna and Sondhi. For the eccentric case, our numerical results give an accurate solution for the eigenvalues and eigenmodes and do not require the uncontrolled simplifying assumptions of previous work. We compare the eigenspectra of the concentric and eccentric drumheads and show that the eccentricity lifts the degeneracy of pairs of concentric eigenmodes. With further refinement, which is part of ongoing work, we believe that our model will find application for numerical sound synthesis of the {\it tabla} and other Indian musical drums. 

The remainder of the paper is organised as follows. In the next section we present our mathematical model for the continuous loading and the boundary value problem that must be solved to obtain the eigenvalues and the eigenfunctions. In Section III we discuss in detail our numerical method, discussing in particular how it leads to a generalised eigenvalue problem. Our results for concentric and eccentric loading are presented in Section IV. We conclude with a summary and discussion of further work.

\section{Mathematical Model}

The drum head of the {\it tabla} is made of leather with the central patch (the {\it sihai}) made of a complex mixture of materials,  as explained in the caption to Fig.~\ref{fig:dayanbayan} The {\it sihai} is approximately eight times as dense as the leather and covers approximately a quarter of the area of the membrane. The {\it sihai} is applied in layers, with each layer made to dry completely before the application of the next layer. This allows a control of the effective mass density of the {\it sihai}. The variation of the harmonicity of the drums with each layer of application of the paste has been studied carefully by Rossing and Sykes\cite{ross82}. 

\begin{figure}
\includegraphics[width = 6cm]{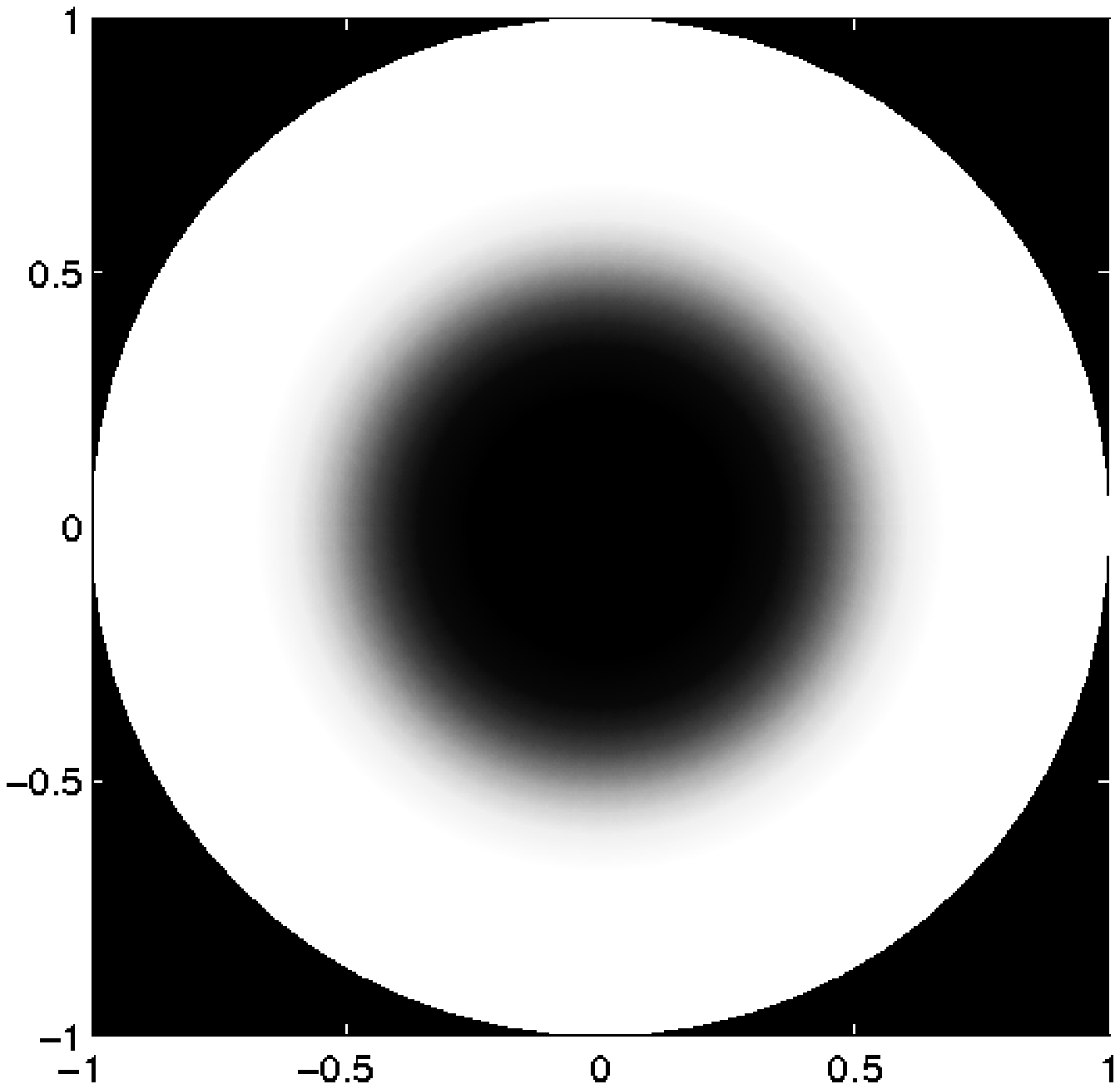}
\caption{\label{fig:rhoprofile}Variation of areal density of the membrane for $\epsilon = 0$. The density is plotted for $\sigma = 2.57$, $k=0.492$ and $\xi=0.091$. We use this form of the density to model the {\it dayan}.}
\end{figure}

It should be clear that the construction of the {\it sihai} is a complex art. However, the most crucial effect of the {\it sihai} is to increase the density in the central region of the drum head. {\it Effectively}, it is possible to think of the drumhead, then, as a membrane of non-uniform density. This forms the basis of our mathematical model. We approximate the {\it tabla} drum head as a circular membrane of unit radius, with a non-uniform areal density $\rho({\bf r})$, where ${\bf r} = (r, \theta)$ is a point on the membrane. Our specific model, which includes both the concentric and eccentric situations, is
\begin{equation}\label{eq:loading}
\rho(r, \theta) = 1 + \frac{\left(\sigma^2 - 1\right)}{2}\left[1 - \tanh\left(\frac{R(r,\theta) - k}{\xi}\right)\right],
\end{equation}
where
\begin{equation}
R(r, \theta) = \sqrt{\left(r\cos\theta - \epsilon\right)^2 + \left(r\sin\theta\right)^2}.
\end{equation}
This function changes smoothly from a value $\rho_2$ at $r=1$ to a value $\rho_1 = \rho_2\sigma^2$ at the center of the loaded region. The change occurs over a region of width $\xi$ along the circle whose equation in polar coordinates is $r = R(r,\theta)$.  For $\epsilon = 0$ the loading is radially symmetric and represents the concentric loading of the ${\it dayan}$. For $\epsilon > 0$, the loading is displaced from the center by a distance $\epsilon$ and then represents the eccentric loading of the ${\it bayan}$. For $\xi \ll 1$, $k^2$ is the ratio of the areas of the loaded to the unloaded regions, requiring $0 < k < 1$. For the concentric case, the variation in density is centrally symmetric, and we display the variation as a function of the radius in Fig.~\ref{fig:rhoprofile}. In the eccentric case, the density depends on both the radius and the polar angle, and Fig.~\ref{fig:eccprofile} illustrates this case. Notice that $\sigma = 1$  corresponds to the uniform membrane, while for fixed $\sigma$, $k$, and $\epsilon$, $\xi\rightarrow 0$ recovers the composite membrane model. Our model, which is smoothly non-uniform, allows a gradual decrease in density of the loaded region and avoids the abrupt change in density of the composite membrane model. Previous attempts at modelling the drumhead by continuous densities have all focussed on the concentric case. Some of these use unphysical models for the mass density\cite{ghosh, rao}, while others need parameters $k$ and $\sigma$ which do not agree with experiment\cite{sid}. 

The equation of motion governing a membrane with spatially varying density $\rho = \rho(r, \theta)$ and uniform tension  $T$ is
\begin{equation}\label{eq:wave}
\rho\ddot u = T\nabla^2 u 
\end{equation}
Here, $u = u(r, \theta, t)$ is the transverse displacement of the membrane at time $t$. For a circular membrane of unit radius clamped at the boundary, the eigenvalue problem is obtained by seeking solutions of 
Eq.\ref{eq:wave} which satisfy the Dirichlet condition 
\begin{equation}\label{eq:clamped}
u(r = 1, \theta, t) = 0
\end{equation} 
The initial-boundary value problem represented by Eq.\ref{eq:wave} and Eq.\ref{eq:clamped} have exact analytical solutions in only a handful of special cases. Of these, the most relevant for the present work are the uniform membrane $\rho(r, \theta) = \rho_0$ and the composite membrane model. To the best of our knowledge, there are no exact analytical solutions when the density varies with both the radius and the angle, as is in the case of Fig.~\ref{fig:eccprofile}. Due to the lack of circular symmetry, the usual strategy of separation of variables in polar coordinates fails, and the eigenfunctions cannot be obtained by a Fourier-Bessel expansion.

This motivates the use of a high-resolution numerical method which we describe in the next section. The advantage of the method, besides its accuracy, is that it solves with equal ease the eigenvalue problem for both the concentric and eccentric cases, facing no difficulty with density variations which depend on both radius and angle. The numerical method also opens the way to a time domain solution for Eq.\ref {eq:wave} which should find application in numerical sound synthesis.

\begin{figure}\label{fig:eccprofile}
\includegraphics[width = 6cm]{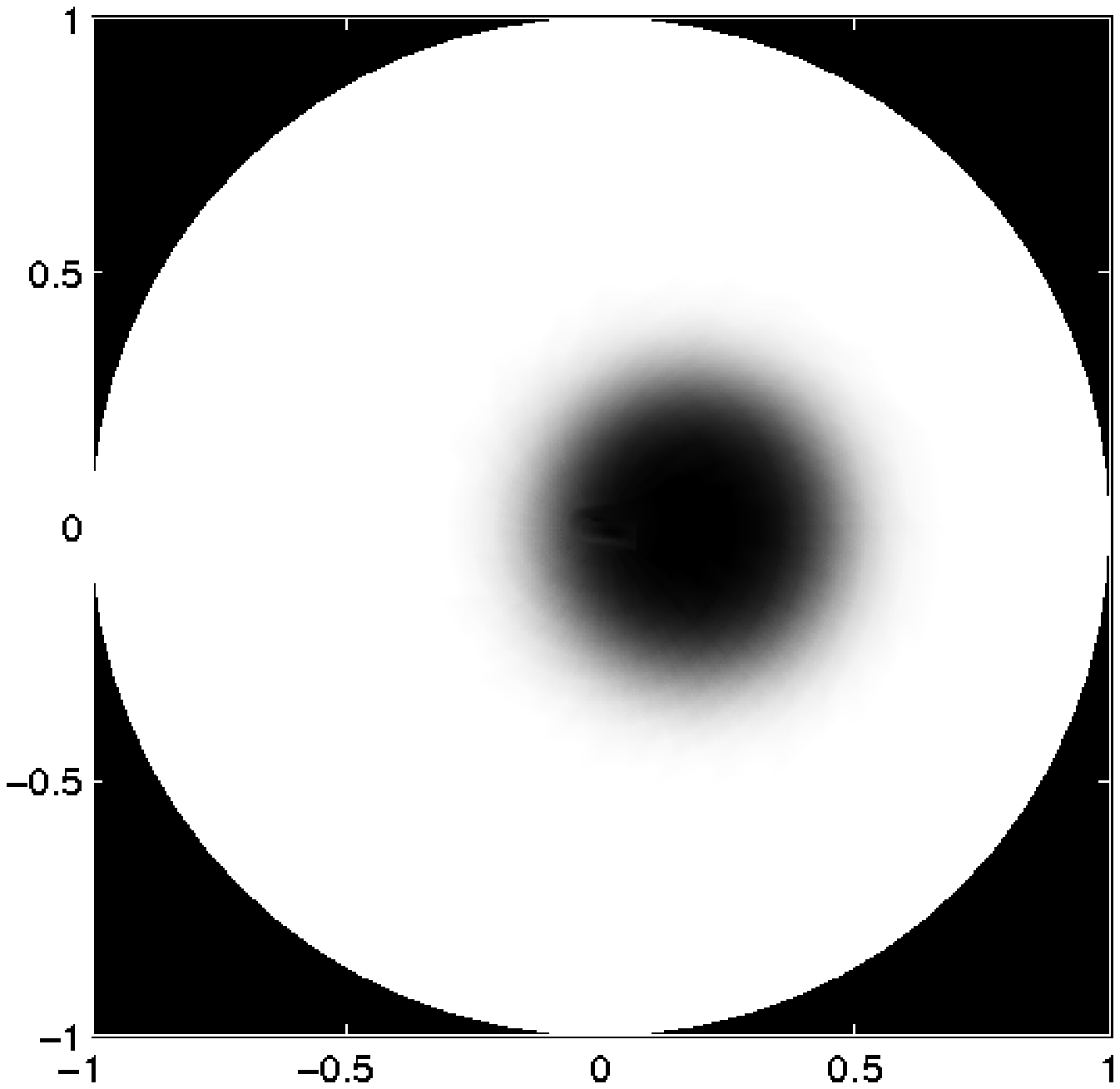}
\caption{\label{fig:eccprofile} Variation of areal density of the membrane for $\epsilon > 0$. The density is plotted for $\epsilon = 0.18$,  $\sigma = 2.57$, $k=0.29$ and $\xi=0.091$. We use this form of the density to model the {\it bayan} }
\end{figure}

\section{Numerical Method}
The eigenvalue equation for the normal modes of the loaded drum is obtained by assuming a solution $u(r, \theta, t) = \Psi_{mn}(r,\theta)\exp(i\omega_{mn} t)$ which transforms Eq.\ref{eq:wave} into
\begin{equation}\label{eq:eig}               
-\omega_{mn}^2\rho(r, \theta)\Psi_{mn}(r, \theta) = T\nabla^2 \Psi_{mn}(r, \theta) 
\end{equation}
Here, $\Psi_{mn}$ is the eigenmode with $m$ nodal lines and $n$ nodal contours. This generalises the labelling used for the uniform circular membrane, where the nodal lines are diameters and the nodal contours are circles. There is an important difference in the mathematical structure of the eigenvalue problem for the uniform and non-uniform membrane. Since the density variation is dependent on position, it cannot be scaled out as in the case of the uniform membrane. This leads to a {\it generalised} eigenvalue problem. The eigenfunctions and eigenvalues are functionals of the non-uniform density $\rho = \rho(r, \theta)$.  

A direct numerical solution of Eq.\ref{eq:eig} is possible using method of varying degrees of accuracy and sophistication. Spectral collocation methods appear to offer the greatest accuracy for the least computational expense for this class of problems. We have therefore used a Fourier-Chebyshev spectral collocation technique, which we describe below, to study the generalised eigenvalue problem in Eq.\ref{eq:eig}.

\begin{figure}
\includegraphics[width = 8cm]{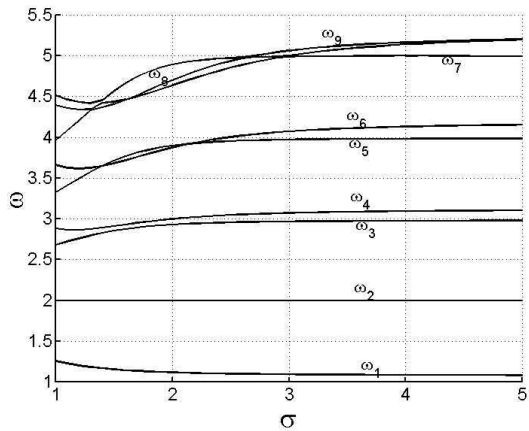}
\caption{\label{fig:eigsigma} Variation of eigenfrequencies with mass density ratio $\sigma$ for the concentric case, $\epsilon =0$, for fixed value of radius ratio $k=0.4$ and fixed smoothness parameter $\xi = 0.091$. The frequencies are normalised by the first overtone. The frequencies are very close being in integer ratios around $\sigma = 3.0$. For large $\sigma$, the frequencies essentially remain constant, but there are several modes that are no longer harmonic.}
\end{figure}

A spectral collocation method proceeds by choosing a set of orthogonal functions and approximating the solution in terms of a linear combination of the orthogonal functions. The approximating function is in the form of an interpolant and matches the solution exactly at a specially chosen set of nodes, the so-called collocation points. The expansions then generate approximations for the derivates which are represented as matrices. Thus, the derivative of A function on a spectral collocation grid is obtained by multiplying the vector of function values by the spectral differentiation matrix. For two or more variables, partial derivatives are obtained by Kronecker products of the differentiation matrices corresponding to each of the independent variables. In Fourier-Chebyshev spectral collocation,
a Fourier expansion is used for the angular coordinate $\theta \in [0, 2\pi]$ and a Chebyshev expansion is used for the radial coordinate $r\in [0, 1]$. The usual Chebyshev expansion is for functions in $[-1, 1]$ and several methods exist for using the Chebyshev expansion for the radial coordinate. Here, we follow the method proposed by Fornberg\cite{fornberg}, using the implementation of Trefethen\cite{trefethen}. The Laplacian in polar coordinates,
\begin{equation}
\nabla^2 = \partial_{r}^2 + r^{-1}\partial_{r} + r^{-2}\partial_{\theta}^2
\end{equation}
is then replaced by the Fourier-Chebyshev differentiation matrix
\begin{equation}
 L = \left(D_1 + RE_1\right)\otimes I_l + \left(D_2 + RE_2\right)\otimes I_r + R^2 \otimes D_{\theta}^{(2)}
\end{equation}
For $N_r$ (odd) Chebyshev collocation points and $N_{\theta}$ (even) Fourier collocation points, the matrices above are representations of the partial derivates on the grid. The two terms with the matrices $D_1$ and $D_2$ represent $\partial_{r}^2$, the two terms with with the matrices $E_1$ and $E_2$ represent $r^{-1}\partial_{r}$ and the last term is the representation of $r^{-2}\partial_{\theta}^2$. $R$ is the diagonal matrix diag($r_j^{-1}$), $1 \leq j \leq (N_r - 1)/2$. The two identity matrices
\begin{equation}
I_l =  \left( \begin{array}{cc}
            I & 0 \\
            0 & I \end{array} \right)
\end{equation}
\begin{equation}
I_r =  \left( \begin{array}{cc}
          0 & I \\
          I & 0 \end{array}\right)
\end{equation}
are formed formed out of the  $N_{\theta/2}\times N_{\theta/2}$ identity matrix $I$. The somewhat complicated looking expression for the Laplacian arises from the use of Fornberg's prescription for handling the radial coordinate using a Chebyshev expansion. Further details are available in Trefethen\cite{trefethen}. The loading function $\rho(r, \theta)$ is itself represented by a matrix ${\mathbf B}$ and the generalized eigenvalue problem can then be formulated as a matrix equation
\begin{equation}
{\mathbf L}\cdot{\Psi} = -\lambda^2{\mathbf B}\cdot\Psi
\end{equation}
We solve for the eigenvectors $\Psi_{mn}$ and the eigenvalues $\lambda_{mn}$ using the Matlab function $\mathtt{eigs}$ which is based on a Cholesky decomposition algorithm. 

We have benchmarked our code with the known eigenvalues of the uniform circular membrane and find spectral convergence with increase in the number of modes. For $N_r = 31$ and $N_{\theta} = 20$ the eigenvalues are accurate to $10$ decimal places. For the non-uniform membrane, higher number of modes are needed, especially in the radial direction, to capture the rapid variation in density for $\xi \ll 1$. The results reported below have $N_r = 65$ and $N_{\theta} = 30$ for the concentric case, and $N_r = 65$ and $N_{\theta} = 56$ for the eccentric case unless stated otherwise. Note that with this choice, the accuracy of the numerical method is several decimal places more than the best reported experimental values. Thus when comparing with experiment,  eigenvalues obtained from the numerical solution can be safely attributed to the model itself and not to numerical errors. 

\begin{table}
\caption{\label{tab:table1} Comparison of eigenfrequencies of the first $9$ eigenmodes of the {\it dayan}, the composite membrane model\cite{rs54}, and the smoothly non-uniform membrane model presented in this work, for $\sigma = 3.125$ and $k=0.4$.  The frequencies are normalised by the first overtone. The figures in parantheses indicate the deviation, in cents, from the experimental value.}
\begin{ruledtabular}
\begin{tabular}{ccll}
Mode&Experimental&Composite&Smooth\\
\hline
$\psi_{01}$&1.03&1.0309 (+1.51)&1.0345 (+7.55) \\
$\psi_{11}$&2.00&2.0000 (0.00)&2.0000 (0.00)\\
$\psi_{21}$&3.00&3.0412 (+23.61)&3.0393 (+22.53)\\
$\psi_{02}$&3.00&3.1546 (+86.99)&3.0534 (+30.54)\\
$\psi_{31}$&4.00&4.0928 (+39.70)&4.0086 (+3.72)\\
$\psi_{12}$&4.00&4.2268 (+95.47)&4.1463 (+62.18)\\
$\psi_{03}$&5.04&4.9794 (-20.94)&4.7784 (-92.27)\\
$\psi_{41}$&5.03&5.1134 (+28.47)&5.0023 (-9.56)\\
$\psi_{22}$&5.08&5.3093 (+76.43)&5.2491 (+56.69)\\
\end{tabular}
\end{ruledtabular}
\end{table}

\section{Results}

We now present our results for the numerical solution of the generalised eigenvalue problem for the smoothly non-uniform membrane. Recall that our model density distribution has four parameters, the ratio of areal densities $\sigma$, the ratio of  radii $k$, the eccentricity parameter $\epsilon$, and the smoothness parameter $\xi$.  We first present results for $\epsilon = 0$, which models the concentric loading of the {\it dayan}, followed by results for $\epsilon > 0$ which models the eccentric loading of the {\it bayan}.

\subsection{Concentric loading}
 In Fig.~\ref{fig:eigsigma} we show the variation of the eigenfrequencies of the first nine eigenmodes as the density contrast is increased, at fixed values of $k$ and $\xi$. All frequencies are normalised by the frequency of the first overtone. The frequency ratios rapidly depart from those of the uniform circular membrane ($\sigma = 1$) to attain harmonic ratios in the neighbourhood of $\sigma = 3$. Our numerical results suggest that for very large $\sigma$ the ratios do not depend on $\sigma$ but have several modes which are no longer harmonic. There is, then, an optimum value of $\sigma$ around $\sigma = 3$ which gives maximally harmonic vibrations. The absolute values of the frequencies decrease monotonically with an increase in the loading, as has been observed previously in analytical and experimental work.

To obtain the values of $\sigma$ and $k$ which produce a maximally harmonic drum, we define a quality function 
$Q(\sigma,k)$  which measures the squared deviation of the frequency of the $i-$th eigenmode from its {\it nearest} integer value.  
\begin{equation}
 Q(\sigma,k) =  \sum_{i= 1}^{N_{max}}(\omega_i(\sigma,k) - h_i)^2
\end{equation}
Here $\omega_i$ denotes the eigenvalue of the $i-$th eigenmode, where $i$ is the rank of the eigenmode when sorted in ascending order of eigenvalue. Eigenvalues from $i=1$ to $i=N_{max}$ are used in calculating the quality. $h_i$ denotes the nearest integer multiple of the fundamental corresponding to frequency $\omega_i$.  Smaller values of $Q(\sigma, k)$ correspond to more harmonic vibrations. 

\begin{figure}
\includegraphics[width = 8cm]{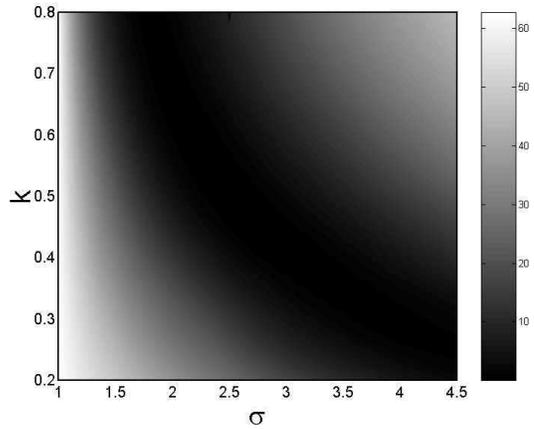}
\caption{\label{fig:qplot} The quality function $Q(\sigma, k)$, defined in the text, as a function of $\sigma$  and $k$. Darker regions correspond to more harmonic vibrations. The most harmonic vibrations are obtained for $\sigma = 2.57$ and $k=0.492$.}
\end{figure}

\begin{figure}
\includegraphics[width = 8cm]{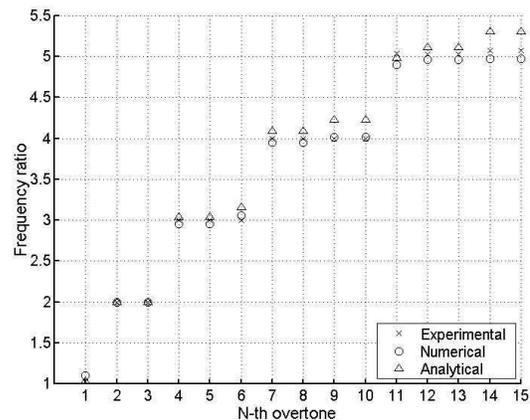}
\caption{\label{fig:eigplot} Comparison of eigenfrequencies of the first $15$ eigenmodes of the {\it dayan}, the composite membrane model\cite{rs54}, and the smoothly non-uniform membrane model presented in this work. The higher overtones are clearly more harmonic when a gradual variation of density of the loaded region is allowed. The parameters used for the smoothly non-uniform membrane are the optimum values of  $\sigma = 2.57$, $k=0.492$, $\xi = 0.091$.}
\end{figure}

\begin{figure*}
\includegraphics[width = 18cm]{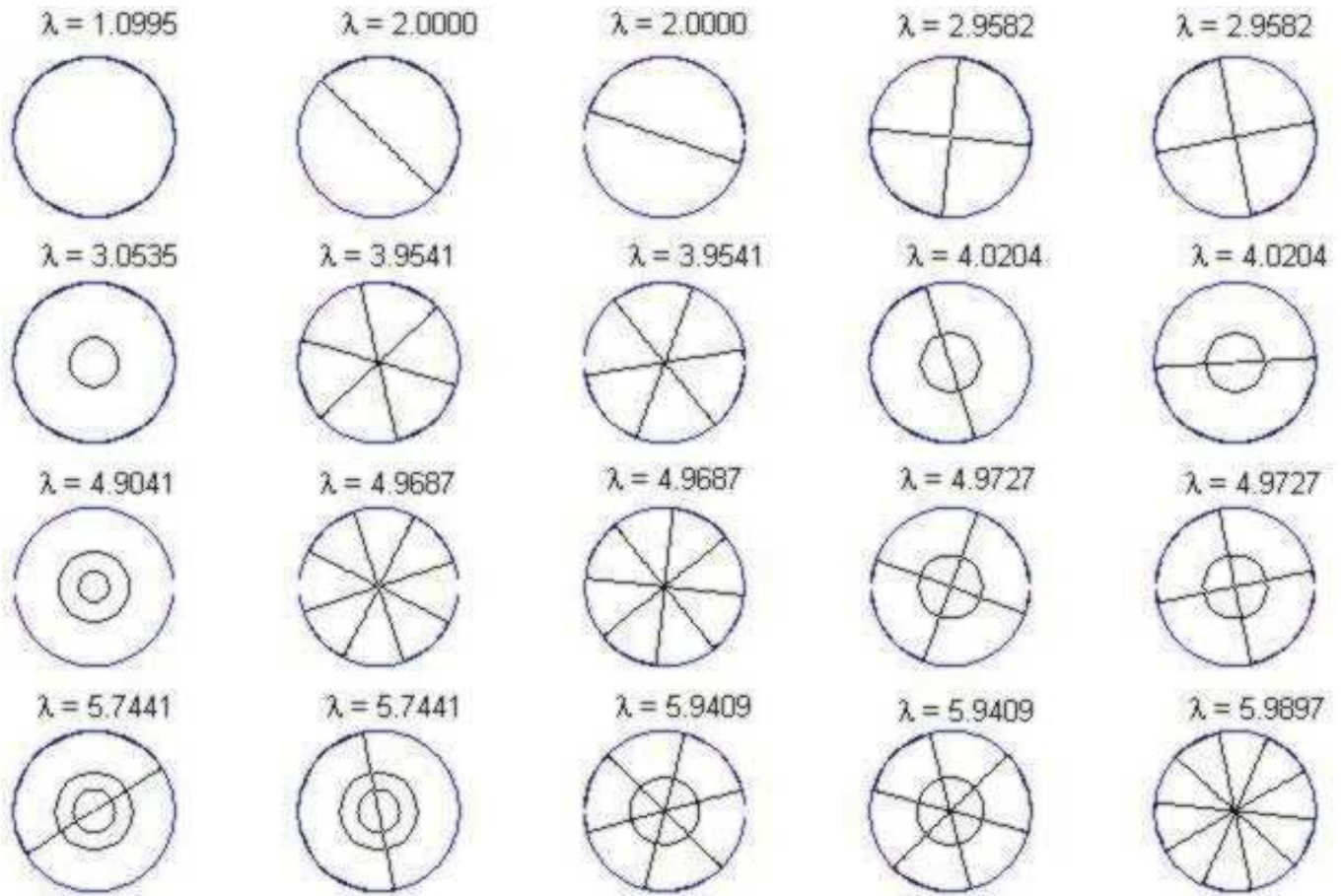}
\caption{\label{fig:conccontours} Nodal contours for the first $20$ eigenmodes of the {\it dayan}. The plots are with $\sigma = 2.57$, $k=0.492$, $\xi = 0.091$.}
\end{figure*}

Thus, to find the best values of $\sigma$ and $k$ at fixed smoothness $\xi$, we are left with a two parameter optimisation problem. We scan the $(\sigma, k)$ parameter space over the range $1\leq\sigma\leq 5$ and $0.2\leq k \leq 0.8$ and obtain the values of $Q(\sigma, k)$, with $N_{max} = 15$. We show the result as a pseudocolour plot in Fig.~\ref{fig:qplot}. In this range, we find that $Q$ has a minimum of $Q_{min} = 0.027$ for $ \sigma_{opt} = 2.57$ and $ k_{opt} = 0.492$.  These values for the density and radii ratios is better when compared with  $Q = 0.074$  for $\sigma = 3.125$, $k=0.4$ for the composite membrane model. Our optimum values, which are obtained without any fitting parameters, fall well within the range of $2.5 < \sigma < 4$, $0.45 < k < 0.55$ that are actually used in the construction of the ${\it dayan}$. Having obtained the optimal values of $\sigma$ and $k$ we further optimise on the smoothness parameter to find a value of $\xi_{opt}$ of $0.091$.

In Table~\ref{tab:table1} we compare the eigenvalues as determined from experiment, from the composite membrane model, and the present model with smooth non-uniformity. We restrict the comparison to the first nine eigenmodes which Raman identified as the being most harmonic. To make a like-for-like comparison with the composite membrane, we use values  $\sigma = 3.125$ and $k=0.4$. Apart from mode $\psi_{03}$, the present model provides a better fit to experimental values than the composite membrane model. 

Going beyond the first nine modes, we find that our model continues to produce harmonic overtones while the composite membrane model shows significant deviations from harmonicity. This is seen clearly in Fig.~\ref{fig:eigplot} where we compare the first $15$ eigenvalues for our model and the composite membrane model, for $\sigma=2.57$ and $k=0.492$. We can, therefore, conclude that the gradual change in density of the loaded region, included here but absent from the composite membrane model, does have an appreciable effect on the musicality of the drum. In Fig.~\ref{fig:conccontours} we show the nodal lines for the first $20$ eigenmodes, including the degenerate ones. These are similar to the nodal lines of an uniform circular membrane.
 
Returning to Fig.~\ref{fig:qplot}, we note that there is a distinct valley of small values in the pseudocolor plot of $Q(\sigma, k)$. This implies that there are many pairs of values of $\sigma$ and $k$ which allow for vibrations that are by and large harmonic. There is, indeed, a wide variation of  $\sigma$ and $k$ in the ${\it dayan}$ from different instrument makers, typically in the range $2.5  < \sigma < 4$ and $0.45 < k < 0.55$. It is tempting to speculate if this could have helped the early makers of the Indian drums, no doubt proceeding by empirical trial-and-error effort, at reaching the optimal values of $\sigma$ and $k$. It should also be mentioned that several Indian musical drums, notably the {\it mridangam},  often use a temporary loading of flour paste (which is applied at the start of the performance and removed afterwards) to ensure harmonic vibrations. It is likely that the principle of central loading was discovered by such temporary application of a heavier material, and later evolved into the more elaborate permanent loading of the {\it sihai}. 

\subsection{Eccentric loading}

\begin{figure}
\includegraphics[width = 8cm]{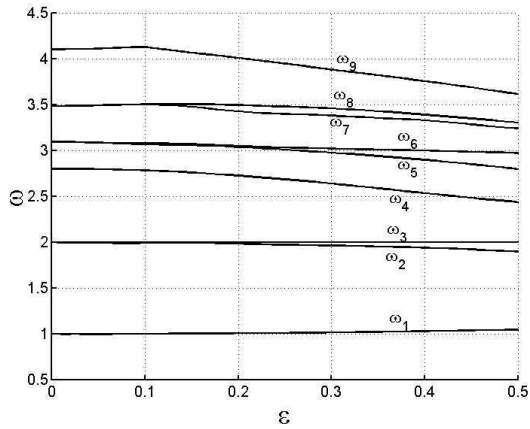}
\caption{\label{fig:eccentriceigs} Variation of eigenvalues with eccentricity for $k=0.29$, $\sigma = 3.125$ and $\xi = 0.091$. The lower eigenvalues remain unchanged for moderate eccentricities.}
\end{figure}

The main function of the eccentric loading in the {\it bayan} is to allow for modulations in the pitch of the drum dynamically, that is, while it is being played. The heel of the hand is moved back and forth along the diameter passing through the centers of the membrane and the eccentrically placed {\it sihai} to modulate the pitch. In this way, the {\it bayan} can produce a distinct sound, not found in any of the other  Indian musical drums. In their experimental measurement \cite{nag91} of the eigenfrequencies of the {\it tabla}, Banerjee and Nag note that only the first few modes are excited by the player's action. The requirement appears to be, then, to allow an eccentric placement of the {\it sihai} and yet retain the harmonicity of the lower modes of vibrations. 

In Fig~\ref{fig:eccentriceigs}, we show how eigenvalues depend on the eccentricity $\epsilon$ at fixed values of $\sigma = 3.125$, $k=0.29$ and $\xi = 0.091$. We see that for  eccentricities up to $0.1$, there is hardly any variation in the eigenspectrum. For larger eccentricity the higher eigenmodes become anharmonic faster than the lower eigenmodes. This is completely consistent with the observation of Nag {\it et al}. 
Our present model thus captures this important feature of the eccentric loading. 

It is worthwhile comparing our numerical results with two prior approximate analytical calculations. In Table~\ref{tab:table2} we compare the eigenvalues as obtained from experiment, an approximate calculation based on the composite membrane model\cite{rs57}, and the present model. The agreement with experimental values is very good, except for eigenmodes which have one nodal circle. We believe that our numerical results are accurate for these modes, but are yet to understand why Ramakrishna's approximate calculation produces better fits to the experimental data. We note that a similar divergence between the model and experiment has been noted in work by Rahman and Sarojini\cite{rah58}, where a variational method was used to calculate the eigenvalues. 

In Table~\ref{tab:table3} we compare our model with the variational calculation of the eigenvalues of the composite membrane. This comparision is of methodological interest only, since the values used do not correspond to the actual values used in the construction of the {\it bayan}.  We note that the agreement is generally not very good, indicating that the variational method possibly overestimates the eigenvalues in this case. 

\begin{figure*}
\includegraphics[width = 18cm]{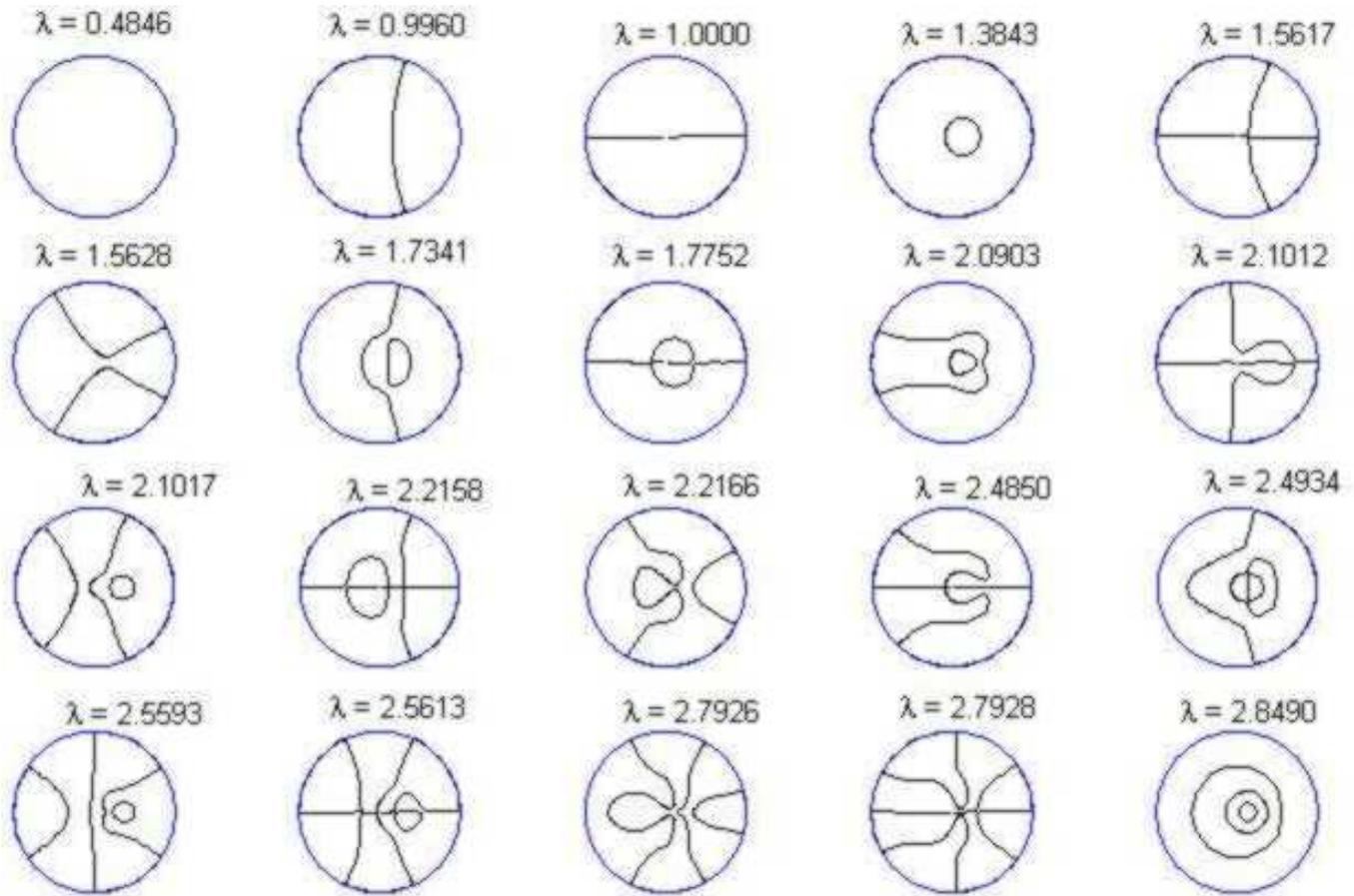}
\caption{\label{fig:ecccontours} Nodal contours for the first $20$ eigenmodes of the {\it bayan}. The plots are with $\sigma = 3.125$, $k=0.29$, $\xi = 0.091$ and $\epsilon = 0.18$.}
\end{figure*}

\begin{figure*}
\includegraphics[width = 14cm]{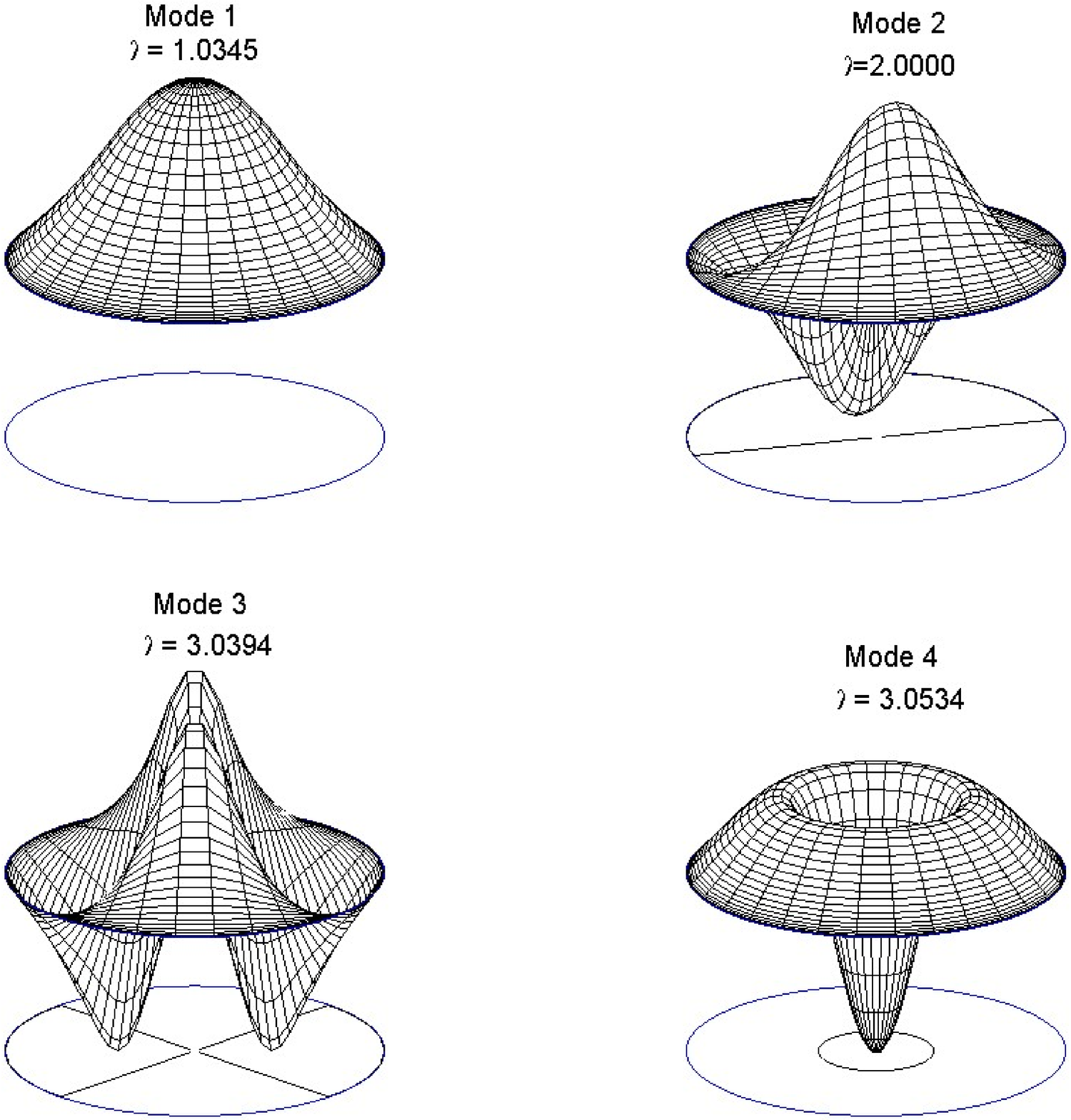}
\caption{ \label{fig:concentricmodes} Eigenmodes for the concentric case }
\end{figure*}

\begin{figure*}
\includegraphics[width = 14cm]{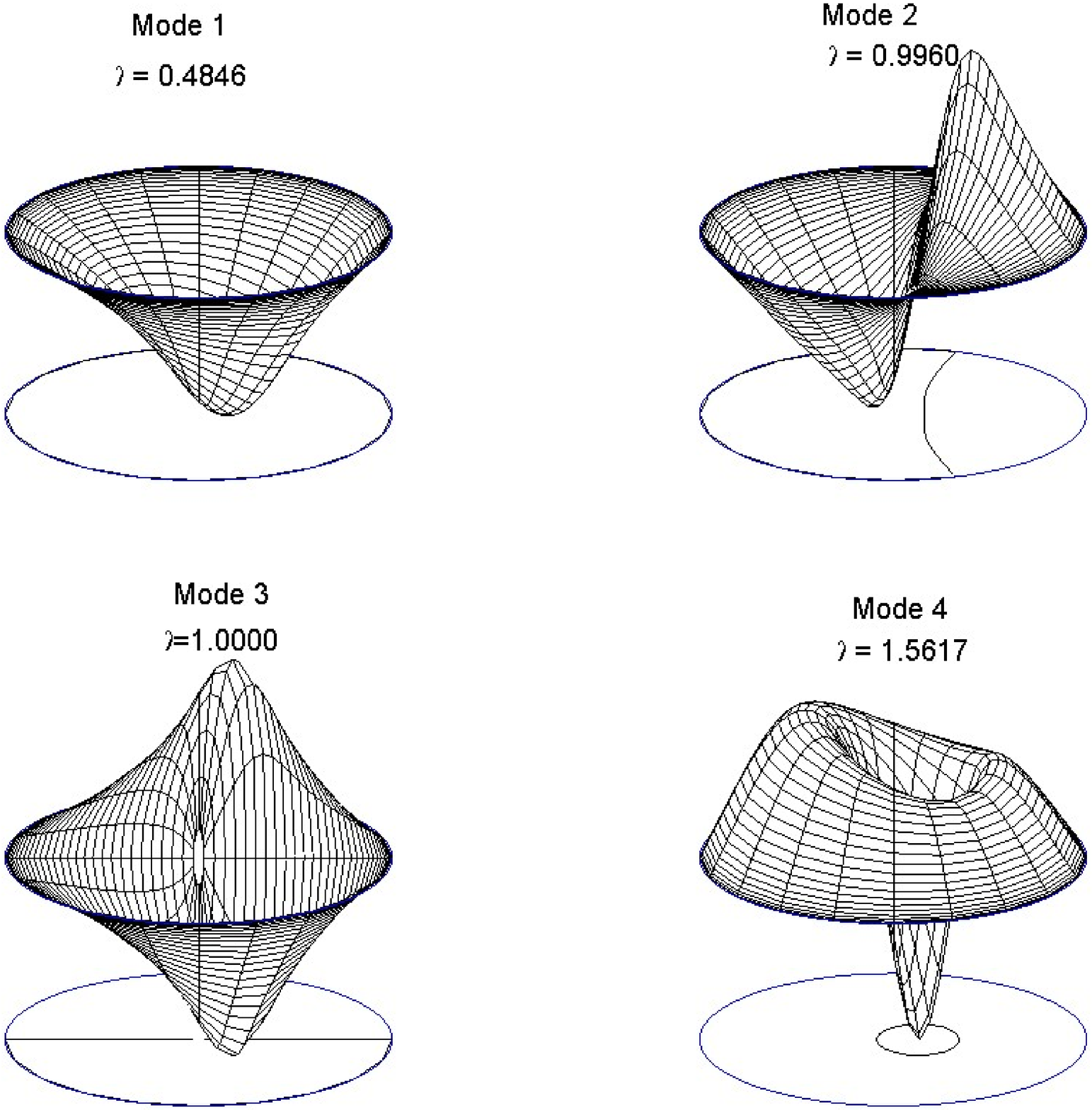}
\caption{\label{fig:eccentricmodes} Eigenmodes for the eccentric case }
\end{figure*}

In Fig.~\ref{fig:ecccontours}, we show the nodal contours of the first $20$ eigenmodes for eccentric loading. It is interesting to see that modes which were degenerate in the concentric case, are no longer degenerate. Further, the nodal diameters now deform into nodal lines which are no longer straight. The nodal circles also deform into closed contours. The lifting of the degeneracies is also evident in Fig.~\ref{fig:eccentriceigs} where we see that each of the lines fork out into two lines at large values of the eccentricity. Finally, in Fig.~\ref{fig:concentricmodes} and Fig.~\ref{fig:eccentricmodes} we compare the first four eigenmodes for concentric and eccentric cases. As should be obvious, the lifting of the degeneracies is clearly seen. Certain eigenmodes no longer have circular symmetry.

\begin{table}
\caption{\label{tab:table2}
Comparison of eigenfrequencies of the first $10$ eigenmodes of the {\it dayan}, the composite membrane model, and the smoothly non-uniform membrane model presented in this work, for $(k=0.29, \epsilon=0.18)$. The figures in parantheses indicate the deviation of frequencies, in cents, from the experimental value.}
\begin{ruledtabular}
\begin{tabular}{ccll}
Mode&Experimental&Analytical& Numerical\\
\hline
$\psi_{01}$&0.54&0.49 (-168.2)&  0.4846 (-187.38)\\
$\psi_{11}$&0.95&0.97 (+36.07)&0.9960 (+81.86)\\
$\psi_{11}$&1.00 &1.00 (0.0)&1.0000 (0.00)\\
$\psi_{21}$&1.52&1.46 (-69.72)&1.5617 (+46.86)\\
$\psi_{21}$&1.54&1.47 (-80.53)& 1.5628 (+25.44)\\
$\psi_{02}$&1.75&1.72 (-29.93)&1.3843 (-405.81)\\
$\psi_{31}$&2.06&1.94 (-103.9)&2.0903 (+25.28)\\
$\psi_{31}$&2.1&1.95 (-128.29)&2.1012(+0.99)\\
$\psi_{12}$&2.32 &2.34 (+14.86)&1.7341 (-503.88)\\
$\psi_{12}$&2.36&2.35 (-7.35)&1.7752(-492.93)\\

\end{tabular}
\end{ruledtabular}
\end{table}

\section{\normalsize Summary}
We have presented a mathematical model, consisting of a membrane of non-uniform density, for the vibrations of the drum head of Indian musical drums. We used a high-resolution numerical method, based on Fourier-Chebyshev collocation, to make an exhaustive study of the variation of the eigenvalues of the model as function of the model parameters. The eigenspectrum of the model agrees very well with the experimentally measured eigenvalues of the {\it tabla}. 

There are several directions in which this present work needs to be extended to make it more realistic. First, we have completely neglected the role of the enclosed air inside the drums. The principal effect of this is to raise the pitch of those modes which, during vibration,  appreciably change the volume of the enclosed air. Modes with nodal circles are strongly affected, the greatest being the fundamental. Modes with nodal diameters are not affected, since the total change in the enclosed volume of air is zero. Preliminary work along these lines has been done by Bhat\cite{bhat91}, but a more systematic numerical study remains to be done. This is part of ongoing work.

Second, our present study focusses only the the real parts of the eigenvalues of the normal modes. We have completely neglected the role of acoustic damping due to the radiation of sound. The damping effects depend quite strongly on the symmetry of the vibrations. For example, the radiation damping of the fundamental, which is in the far field is an acoustic monopole, is quite different from that of the first overtone, which in the far field is an acoustic dipole. We have seen no numerical study of the radiation damping problem for the Indian drums, though extensive work has been done for the uniform circular membrane. This is a problem for further study. 

Third, with further refinement our model and the numerical method should find application in numerical sound synthesis. With the increasing power of computer hardware, it is now possible to simulate in real time, physical models, albeit simple ones, of musical elements like strings, membranes, and plates. Numerical sound synthesis will take advantage of growing computational power and we believe that it will be possible to have realistic numerical models of the Indian musical drums using the model and numerical method presented here.

\begin{table}
\caption{\label{tab:table3} A comparison of the variational\cite{rah58} and numerical eigenvalues for eccentric loading,$(k=0.4, \epsilon=0.18)$.}
\begin{ruledtabular}
\begin{tabular}{ccc}
Mode&Variational& Numerical\\
\hline
$\psi_{01}$&1&1\\
$\psi_{11}$&1.9&1.9199\\
$\psi_{11}$&1.96&1.9215\\
$\psi_{02}$&3.08&2.9026\\
$\psi_{21}$&2.98&2.9207\\
$\psi_{21}$&2.98&2.9215\\
$\psi_{12}$&4.04&3.7715\\
$\psi_{12}$&4.15&3.8323\\
\end{tabular}
\end{ruledtabular}
\end{table}

Fourth, we note that our model of the non-uniform membrane is simplified. The {\it sihai}, as mentioned earlier, is actually a complex material, made of several ingredients like soot, iron filings, flour and other polymerising substances. The making and application of the {\it sihai} is an art, and we believe our smoothly non-uniform membrane model captures the subtle physics of the {\it sihai} in a gross manner. There remains considerable scope for improvement of mathematical models of the  drum head.

Finally, we end with an interesting speculation. Marc Kac \cite{kac} posed the isospectral problem for a two-dimensional Laplacian by wittily asking ``Can one hear the shape of a drum ?''. Very recently it has been shown that one cannot hear the shape of a drum \cite{gordon91}, that is, there exist distinct boundaries in which the Laplacian operator with Dirichlet boundary conditions has the same spectrum. One may now ask, is the same true for a non-uniform membrane ? In other words, ``Can one hear the shape of an Indian drum ?''.

\begin{acknowledgements} RA wishes to thank Dhananjay Modak for useful discussions regarding the construction of the {\it tabla}. We thank  
deodesign.wordpress.com for permission to reproduce Figs.~\ref{fig:tablatopview} and \ref{fig:dayanbayan}. This work was funded in part by grant EPSRC GR/S10377 at the University of Edinburgh. 
\end{acknowledgements}

\bibliography{spectraldrums}
\end{document}